\documentclass[a4paper]{article}

\usepackage{INTERSPEECH2021}
\usepackage{multirow}

\title{Optimising Hearing Aid Fittings for Speech in Noise with a Differentiable Hearing Loss Model}
\name{Zehai Tu, Ning Ma, Jon Barker}
\address{University of Sheffield, Department of Computer Science, Sheffield, UK}
\email{\{ztu3, n.ma, j.p.barker\}@sheffield.ac.uk}

\begin{document}

\maketitle
\begin{abstract}
% Current hearing aids normally provide amplification based on a general prescriptive fitting together with noise suppression. However, they still fail to perform well for speech in noise because of the imperfection of noise suppression and the lack of considering the effects of both environmental disturbances and the noise suppression algorithm when developing the fittings. Techniques which can provide adaptive fitting development are thus needed. This paper provides such a method by taking advantage of recent successful data-driven machine learning techniques. A differentiable hearing loss model is proposed and used to optimise fittings with data of speech in different noise classes and processed with or without noise suppression. The objective evaluation not only shows the advantages of such customised fittings but also challenges the necessity of a conventional noise suppression in hearing aids. 

Current hearing aids normally provide amplification based on a general prescriptive fitting, and the benefits provided by the hearing aids vary among different listening environments despite the inclusion of noise suppression feature. Motivated by this fact, this paper proposes a data-driven machine learning technique to develop hearing aid fittings that are customised to speech in different noisy environments. A differentiable hearing loss model is proposed and used to optimise fittings with back-propagation. The customisation is reflected on the data of speech in different noise with also the consideration of noise suppression. The objective evaluation shows the advantages of optimised custom fittings over general prescriptive fittings.

\end{abstract}

\section{Introduction}

% The problem and our solution in brief

The most common complaint of hearing aid users is that they struggle to understand speech in noisy situations~\cite{lesica2018hearing, brons2015acoustical}. This is despite the fact that hearing aids are able to provide sufficient amplification, and despite the fact that modern hearing aids often include noise suppression algorithms. Ultimately better source separation algorithms might solve this problem, but in this paper we investigate whether speech intelligibility in noise can be improved by data-driven approaches to parameter-tuning in current hearing aid designs. In particular we look at the potential for replacing traditional hearing aid fitting formulae with scene-dependent fitting algorithms.

% The current approach

Hearing aids are currently tuned for individuals using one of various prescriptive fitting formulae. These formulae have similar design goals which link the listener's audiogram (frequency dependent hearing thresholds) to a prescribed degree of frequency dependent amplification. 
% They have simple design goals. 
The National Acoustic Laboratories' Revised (NAL-R) fitting~\cite{byrne1986national} aims to achieve equal loudness across all frequency bands in linear hearing aids. The more recent NAL-NL1 and NAL-NL2~\cite{keidser2011nal} were designed to work with hearing aids that employ non-linear dynamic range compression algorithms and are designed to theoretically maximise speech intelligibility while not making the signal louder than usual. They achieve this using a loudness model~\cite{moore1997model} and the speech intelligibility index (SII)~\cite{ansi1997s3}. Other fitting formulae include CAMEQ and CAMEQ2-HF~\cite{moore1999use, moore2010development} and the widely used DSLv5~\cite{scollie2005desired}.

% Why the current approach is not optimal

The fitting formulae approach to hearing aid gain setting is remarkably successful given that a single formulae is used to cover all listening conditions. However, the question naturally arises as to whether a better result could be achieved by using noise-dependent fittings, and if so, how should these fittings be optimised? This is particularly relevant now that environment classification algorithms are available to automatically detect whether a user is, say, in a domestic living room, in a noisy cafe or standing by a busy road intersection, and hence good be used to switch gain settings. It should also be considered that hearing aids now apply increasingly sophisticated (but imperfect) noise-reduction algorithms (e.g., adaptive filtering ~\cite{vary2006digital}, spectral subtraction\cite{boll1979suppression, bentler2006digital}, spatial filtering ~\cite{levitt2001noise}) that can altered the signal in ways that have not been considered in the design of NAL-NL2 etc. Further, recent hearing aids are using environmental classification algorithms~\cite{lamarche2010adaptive, nordqvist2004efficient} to allow the characteristics of the noise suppression algorithms to be tuned separately for different noise types~\cite{bentler2006digital}. This further complicates the requirements of a fitting formulae. Logically, hearing aid gains should be optimised in consideration of the perception of the \textit{processed} noise-reduced signal that the hearing aid delivers.

% How we can do better using data-driven techniques

This study explores the possibilities of improving current hearing aid performance by developing fittings specific to different listening environments and noise-reduction processing. This is achieved using a model of hearing loss and a data-driven machine learning method which, inspired by the recent success of deep learning techniques, is optimised with the back-propagation algorithm.  Our previous work using a similar data-driven fitting optimisation framework, DHASP~\cite{tu2021dhasp}, has shown that hearing aid fittings optimised by a data-driven approach can achieve high intelligibility for speech in quiet conditions. This work extends the DHASP framework to speech in noise, and proposes a new differentiable hearing loss model for the optimisation. The hearing loss model is approximated to the Cambridge Auditory Group MSBG model~\cite{baer1993effects, baer1994effects, moore1993simulation, stone1999tolerable} from the recent released Clarity Challenge~\cite{graetzer2020clarity}. For evaluation, various objective evaluation metrics were used and the performances of the optimised general fittings and custom fittings along with an open-source well-recognised prescriptive fitting are compared across a variety of noise types and hearing abilities. The results show the advantages of the optimised custom fittings, suggesting the proposed method could bring potential improvements for current hearing aids.

This paper is organised as follows. 
%In Section~\ref{sec:background}, background on the development of hearing aid prescriptive fittings and noise suppression is introduced. 
Section ~\ref{sec:method} presents the proposed fitting optimisation method. Section~\ref{sec:experiments} describes the evaluation metrics, the databases and the experimental setup. The results comparing fittings in three contrasting noise environments, with and without noise reduction are presented and discussed in Section~\ref{sec:results}. Section~\ref{sec:conclusions} concludes the paper and presents future directions.

\section{Method}
\label{sec:method}
\begin{figure}[t]
  \centering
  \includegraphics[width=\linewidth]{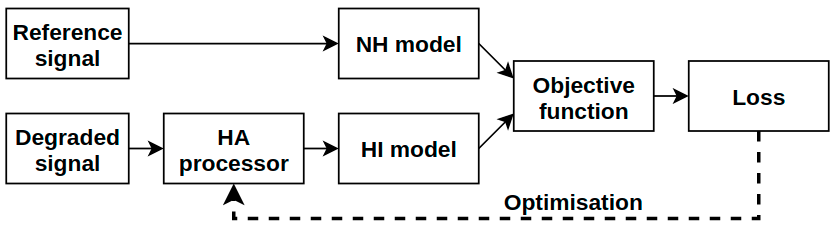}
  \caption{Overall workflow}
  \label{fig:overall_structure}
  \vspace{-.4cm}
\end{figure}

The overall workflow of the proposed method is shown in Fig.~\ref{fig:overall_structure}. The degraded signal represents a noisy speech signal processed with or without a noise suppression algorithm. To simulate the typical signal pathway of a hearing aid (HA) user, the degraded signal is enhanced by a HA processor before being processed by a hearing impaired (HI) model. Its difference to a reference signal processed by a normal hearing (NH) model is measured and used as loss to optimise the HA processor with back-propagation. The NH and HI models are represented by a hearing loss model, whose characteristics are based on a listener's audiogram. Both the HA processor and the hearing loss models operate at a 44.1\,kHz sampling rate, and all signals are presented at 65 dB as the sound pressure level (SPL) of normal conversation. A high-performance deep learning library PyTorch~\cite{paszke2019pytorch} is used for the implementation to retrieve the gradients for the optimisation.

\subsection{Hearing-aid processor}
As the SPL of input speech is constant at 65 dB, a finite impulse response (FIR) filter is used as the HA processor providing constant amplification excluding dynamic-range compression features. The amplification provided depends on six trainable parameters, which represent the insertion gains at [250, 500, 1000, 2000, 4000, 6000] Hz to be consistent with the audiograms. The frequency response is then obtained with linear interpolation, and iFFT is applied to retrieve the impulse response. A Hann window is subsequently multiplied with the impulse response.

\subsection{Hearing loss model}
\begin{figure}[t]
  \centering
  \includegraphics[width=\linewidth]{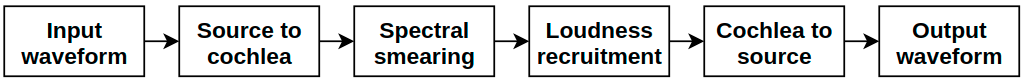}
  \caption{Structure of the hearing loss model}
  \label{fig:hearinglossmodel_structure}
  \vspace{-.4cm}
\end{figure}

The MSBG model can be considered as a hearing impairment simulator, which consists of the simulations of acoustic transformation between sound source and cochlea, spectral smearing, and loudness recruitment. The structure of the model is shown in Fig.~\ref{fig:hearinglossmodel_structure}. The hearing loss model proposed in this work is a differentiable approximation to the MSBG model, and the differences are in filter implementation and envelope retrieval. All infinite impulse response (IIR) filters in the MSBG model are approximated with FIR filters, and Hilbert transformation is used to extract the envelopes, so that the computation can be performed in parallel for fast optimisation using GPUs.

\subsubsection{Source to cochlea transformation}
The transformation of the sound pressure level from a sound source to the cochlea is derived from the combination of a free field and a middle ear transfer function. The free field transfer function~\cite{shaw1974transformation} approximates the acoustic changes during sound propagation from the free field to the eardrum. The middle ear transfer function~\cite{killion1978revised} simulates the acoustic alterations of sound in the middle- and inner-ear before arriving at the cochlea. The overall transformation is implemented using an FIR filter, whose frequency response is the combination of the free field and the middle ear frequency-gain tables.

\subsubsection{Spectral smearing}
Spectral smearing~\cite{baer1993effects} is used in the hearing loss model to simulate reduced frequency selectivity, which is one of the major deficits in the sound analysis ability of cochlear hearing loss. Experimental results showed that this technique leads to little effect on the intelligibility of speech in quite, but a large effect on speech in noise~\cite{baer1993effects} or interfering speech~\cite{baer1994effects}. This is generally consistent with the phenomenon that impaired frequency selectivity contributes largely to the difficulty of understanding speech in noise for listeners with cochlea hearing loss.

Input waveform signals are first processed by STFT with Hamming windows. Smearing is then performed to the power spectrogram, and the phase remains unchanged for iSTFT after smearing. Given the input spectrogram $X$ and the output spectrogram $Y$, the spectrum smearing function is expressed as:
\begin{equation}
    Y = A_{N}^{-1}A_{W}X,     
\end{equation}
where $A_{N}$ and $A_{W}$ are the matrices representing the normal and the widened auditory filterbanks, respectively. For each auditory filter within the filterbank, the form is given by:
\begin{equation}
    W(g) = (1 + pg)\exp(-pg),
\end{equation}
where $W(g)$ is the intensity weighting function describing the filter shape in the frequency domain, $g$ is the frequency difference from the center frequency $f_{c}$ divided by $f_{c}$, and $p$ is the parameter determining the sharpness of the auditory filter~\cite{moore1983suggested}. The value of $p$ is computed as:
\begin{equation}
    p = \frac{4f_{c}}{r\times\text{ERB}},
\end{equation}
where ERB is the equivalent rectangular bandwidth~\cite{glasberg1990derivation} calculated as $24.7\times(0.00437f_{c}+1)$, and the widening factor $r$ differs for the lower and upper sides of the filter, denoted as $r_{l}$ and $r_{u}$, respectively. The values of $r_{l}$ and $r_{u}$ are dependent on the degree of hearing loss. Each auditory filter is at last calibrated by dividing $24.7\times \text{ERB}(r_{l} + r_{u})/2$ to remove an upward tilt in the excitation pattern, which is caused by the increasing of the bandwidth as the $f_{c}$ grows.

\subsubsection{Loudness recruitment}
It is observed that the response of a damaged cochlea to low-level sounds is much smaller than a normal one, while the response to high-level sounds is roughly the same as normal~\cite{moore1993simulation}. This is simulated by a dynamic range compression mechanism.

A group of Gammatone filters are firstly used to extract the fine structures $x(n)$ of the smeared waveform signal. The $i^{th}$ filter $h^{(i)}$ of filterbank is expressed as: 
\begin{equation}
    h^{(i)}(t) = A^{(i)} t^{\left(N^{(i)}-1\right)} e^{-2 \pi b^{(i)} t}\cos \left(2 \pi f^{(i)} t\right),
\end{equation}
where $f^{(i)}$ is the centre frequency, $b^{(i)}$ is the bandwidth computed as $1.019\times\text{ERB}$, $N$ is the order of the Gammatone filter ($4$ in this study), and $A^{(i)}$ is the amplitude to normalise the filters. The number of filters $I$ differs according to the hearing abilities. The bandwidths of the filters are broadened two or three times for moderate or moderate to severe hearing loss, respectively. The outputs of the filters are aligned in the time domain to ensure the peaks for all channels are coincident for a pulse input. This alignment will make the mixture of the outputs of the filters generally sound almost identical to the input signal. The envelope $E(n)$ of each channel is retrieved with Hilbert transformation followed by a group of low pass filters for smoothing. The waveform output signal $y(n)$ is then recruited as:
\begin{equation}
    y(n) = \sum_{i=1}^{I}\left(\frac{E^{\left(i\right)}\left(t\right)}{E_{\theta}}\right)^{\left(\frac{\theta}{\theta - \text{HL}^{\left(i\right)}}-1\right)}x^{(i)}\left(n\right),
\end{equation}
where HL is the audiometric hearing loss in dB, $\theta$ is the maximal loudness threshold which is set 105\,dB, and $E_{\theta}$ is the corresponding envelope magnitude.

\subsubsection{Cochlea-to-source transformation}
The recruited signal is lastly processed by a cochlea-to-source transfer FIR filter, whose frequency response is the additive inverse in dB of the frequency response of the source-to-cochlea transformation filter.

\subsection{Objective function}
Given the simulated reference signal $y_{r}(n)$, i.e. the clean normal hearing signal, and the processed hearing impaired signal $y_{p}(n)$, STFT is firstly applied to retrieve the corresponding spectrograms $Y_{r}(m, k)$ and $Y_{p}(m, k)$. The objective function consists of a spectrogram reconstruction loss $L_{spec}$ and a sound pressure level loss $L_{spl}$.
$L_{spec}$ is expressed as:
\begin{equation}
    L_{spec} = 20\log_{10}\left(\frac{1}{mk}\sum_{m, k}\left|Y_{p}\left(m, k\right) - Y_{r}\left(m, k\right)\right|\right),
\end{equation}
and $L_{spl}$ is computed as:
\begin{equation}
    L_{spl} = 20\log_{10}\left(\sqrt{\frac{1}{n}\sum_{n}y_{p}(n)^{2}} - \sqrt{\frac{1}{n}\sum_{n}y_{r}(n)^{2}}\right).
\end{equation}
The overall objective function is expressed as:
\begin{equation}
    L = 
    \begin{cases}
    L_{spec} + \alpha L_{spl},& \text{if}\ L_{spl}\geq 0\\
    L_{spec},              & \text{otherwise}
\end{cases},
\label{eq8}
\end{equation}
where $\alpha$ is a weighting coefficient. $L_{spl}$ is used to prevent over-amplification, which could lead to listening discomfort for the listeners, and thus it is not applied if negative.

\section{Experiments}
\label{sec:experiments}
\subsection{Evaluation}
The hearing-aid speech perception index (HASPI)~\cite{kates2014hearinghaspi} and hearing-aid speech quality index (HASQI)~\cite{kates2014hearinghasqi}, which are developed based on a physiological auditory model, are used to evaluate the speech intelligibility and quality, respectively. Segmental frequency weighted signal-to-noise (FWSNR) ratio is also used for evaluation, and the implementation  is from~\cite{loizou2013speech}. The FWSNRs are measured after the amplified signals processed by the MSBG model. 

The open-source NAL-R prescription is used as the baseline prescription, which is widely recognised and more importantly tested in subjective experiments. To be consistent with the HA processor, the derivation of the NAL-R frequency response is based on the hearing losses at [250,  500,  1000,  2000,  4000,  6000] Hz. The Wiener filtering algorithm, popular for noise suppression in hearing aids, is used in this work as the noise suppression front end. Fittings optimised on clean data are regarded the general fittings (G). Meanwhile, the custom fittings include the fittings optimised on noisy data (Cn) and the fittings optimised on the Wiener filtering enhanced noisy data (Cw).

\subsection{Databases}
Three standard audiograms N1, N2, and N4 from~\cite{bisgaard2010standard} are used in this study, which represent three hearing loss categories: mild (N1), moderate (N2), and moderate to severe (N4). Their hearing losses at different frequencies are shown in Table~\ref{tab:audiogram}.
%The hearing losses of N1, N2, and N4 at [250,  500,  1000,  2000,  4000,  6000] Hz are [10, 10, 10, 15, 30, 40], [20, 20, 25, 35, 45, 50], and [55, 55, 55, 65, 75, 80] dB, respectively.

\begin{table}[t]
  \caption{Hearing losses of the audiograms used in this study.}
  \vspace{-.2cm}
  \label{tab:audiogram}
  \centering
  \resizebox{\linewidth}{!}{
  \begin{tabular}{l | l l l l l l }
    \toprule
    & 250\,Hz&  500\,Hz&  1\,kHz&  2\,kHz&  4\,kHz&  6\,kHz\\
    \midrule
    N1 & 10\,dB & 10\,dB & 10\,dB & 15\,dB & 30\,dB &40\,dB \\
    N2 & 20\,dB & 20\,dB &25\,dB &35\,dB &45\,dB &50\,dB \\
    N4 & 55\,dB &55\,dB &55\,dB &65\,dB &75\,dB &80\,dB \\
    \bottomrule
  \end{tabular}}
  \vspace{-.4cm}
\end{table}

The noisy speech corpus from~\cite{valentini2016investigating} is used, which was mixed using speech utterances from the the Voice Bank Corpus~\cite{veaux2013voice}, and noises from the recordings of the first channel of the Demand database~\cite{thiemann2013demand}. A set of 56 speakers is used for training, and another set of 28 speakers is divided into the validation and test sets. There are around 400 utterances from each speaker. 10 types of noises are mixed with the utterances, at the SNRs of 0, 5, 10, and 15\,dB according to ITU-T P.56 standard~\cite{itu1993objective}. All signals are filtered with a high pass filter whose cut-off frequency is 80 Hz to eliminate non-speech interference.
% In the experiments, three categories of noises are selected, including (1) traffic noise which is comparatively stationary and mostly in low frequency, (2) kitchen noise which is less stationary and mostly in high frequency, and (3) 6-speaker babble noise which is non-stationary and the spectrum is similar to the clean speech.
Three categories of noises are selected in the experiments: traffic, kitchen, and babble. The traffic noise and kitchen noise are comparatively more stationary, and mainly distributed in low and high frequencies, respectively. The babble noise is less stationary, and its spectrum overlaps clean speech.

\begin{table*}[t!]
  \caption{Evaluation scores of various fittings applied to noisy speech before and after the enhancement of Wiener filtering. N: NAL-R prescriptive fittings, G: optimised general fittings, +W: using enhanced noisy speech by Wiener filtering, Cn: custom fittings optimised on noisy data, Cw+W: custom fittings optimised on Wiener filtering enhanced noisy data with Wiener filtering. The single best score in each group is indicated in bold.}
  \label{tab:scores}
  \vspace{-.0cm}
  \centering
    \resizebox{\textwidth}{!}{
  \begin{tabular}{c c | c c c c c c | c c c c c c | c c c c c c}
    \toprule
    & & \multicolumn{6}{c}{\textbf{Traffic}} & \multicolumn{6}{c}{\textbf{Kitchen}} & \multicolumn{6}{c}{\textbf{Babble}}\\
    & & N & N+W & G & G+W & Cn & Cw+W & N & N+W & G & G+W & Cn & Cw+W & N & N+W & G & G+W & Cn & Cw+W \\
    \toprule
    \multirow{3}{*}{\textbf{N1}}
    & HASPI & 0.92 & 0.93 & 0.92 & 0.95 & 0.93 & 0.95 & 0.99 & 0.98 & 0.99 & 0.99 & 0.99 & 0.99 & 0.68 & 0.65 & 0.66 & 0.67 & 0.68 & 0.67\\
    & HASQI & 0.25 & 0.23 & 0.26 & 0.27 & \textbf{0.28} & 0.26 & 0.42 & 0.30 & 0.42 & 0.35 & \textbf{0.44} & 0.35 & 0.15 & 0.13 & 0.15 & 0.14 & \textbf{0.16} & 0.14\\
    & FWSNR & 5.40 & 6.08 & 5.82 & 6.40 & \textbf{7.59} & 6.22 & 9.54 & 8.77 & 7.42 & 8.65 & \textbf{10.71} & 8.76 & 4.26 & 4.21 & 4.41 & 4.36 & \textbf{6.00} & 4.23\\
    \midrule
    \multirow{3}{*}{\textbf{N2}}
    & HASPI & 0.87 & 0.81 & 0.88 & 0.91 & 0.89 & 0.91 & 0.97 & 0.91 & 0.98 & 0.98 & 0.98 & 0.98 & 0.57 & 0.50 & 0.59 & 0.59 & \textbf{0.60} & 0.59\\
    & HASQI & 0.18 & 0.13 & 0.20 & 0.19 & \textbf{0.21} & 0.19 & 0.23 & 0.17 & 0.30 & 0.24 & \textbf{0.30} & 0.24 & 0.11 & 0.08 & 0.12 & 0.10 & 0.12 & 0.10\\
    & FWSNR & 5.63 & 6.00 & 7.08 & 6.31 & \textbf{7.61} & 6.30 & \textbf{9.64} & 8.54 & 7.84 & 7.77 & 8.39 & 7.65 & 4.35 & 4.07 & 5.38 & 4.35 & \textbf{5.48} & 4.24\\
    \midrule
    \multirow{3}{*}{\textbf{N4}}
    & HASPI & 0.21 & 0.16 & 0.52 & 0.40 & 0.52 & 0.39 & 0.23 & 0.20 & 0.61 & 0.53 & 0.61 & 0.51 & 0.13 & 0.09 & \textbf{0.31} & 0.24 & 0.30 & 0.23\\
    & HASQI & 0.03 & 0.02 & 0.08 & 0.05 & 0.08 & 0.05 & 0.03 & 0.03 & 0.08 & 0.07 & 0.08 & 0.07 & 0.02 & 0.02 & \textbf{0.06} & 0.04 & 0.05 & 0.04\\
    & FWSNR & 0.91 & 2.06 & 5.12 & 4.44 & \textbf{5.24} & 4.40 & 1.86 & 2.21 & 6.55 & 5.71 & \textbf{6.63} & 5.65 & 0.37 & 1.01 & 3.52 & 2.81 & \textbf{3.66} & 2.80\\
    \bottomrule
  \end{tabular}}
\end{table*}

\subsection{Experimental setup}
The Wiener filter implementation is based on the method proposed in~\cite{plapous2006improved}. 
% The filtered signals are normalised to the same sound pressure level as the input noisy signals.
% All utterances are resampled to 44.1kHz and the presentation level is set as 65 dB.
The parameter setting in the differentiable hearing loss model is consistent with the MSBG model. In the spectral smearing, [$r_{l}$, $r_{u}$] are set as [4.0, 2.0], [2.4, 1.6], and [1.6, 1.1] for the mild, moderate, and moderate to severe hearing loss, respectively. In the loudness recruitment, the Gammatone filterbank consists of 36, 28, or 19 filters respectively for the three types of hearing loss. The HA processors are trained with a batch size of 128 for 500 epochs using the Adam optimiser with a learning rate of 1$e$-2. The parameters are initialised as the NAL-R fitting. The weighting coefficient in Eq.~\ref{eq8} is set to 5.

\begin{figure}[thb]
  \centering
  \includegraphics[width=\linewidth]{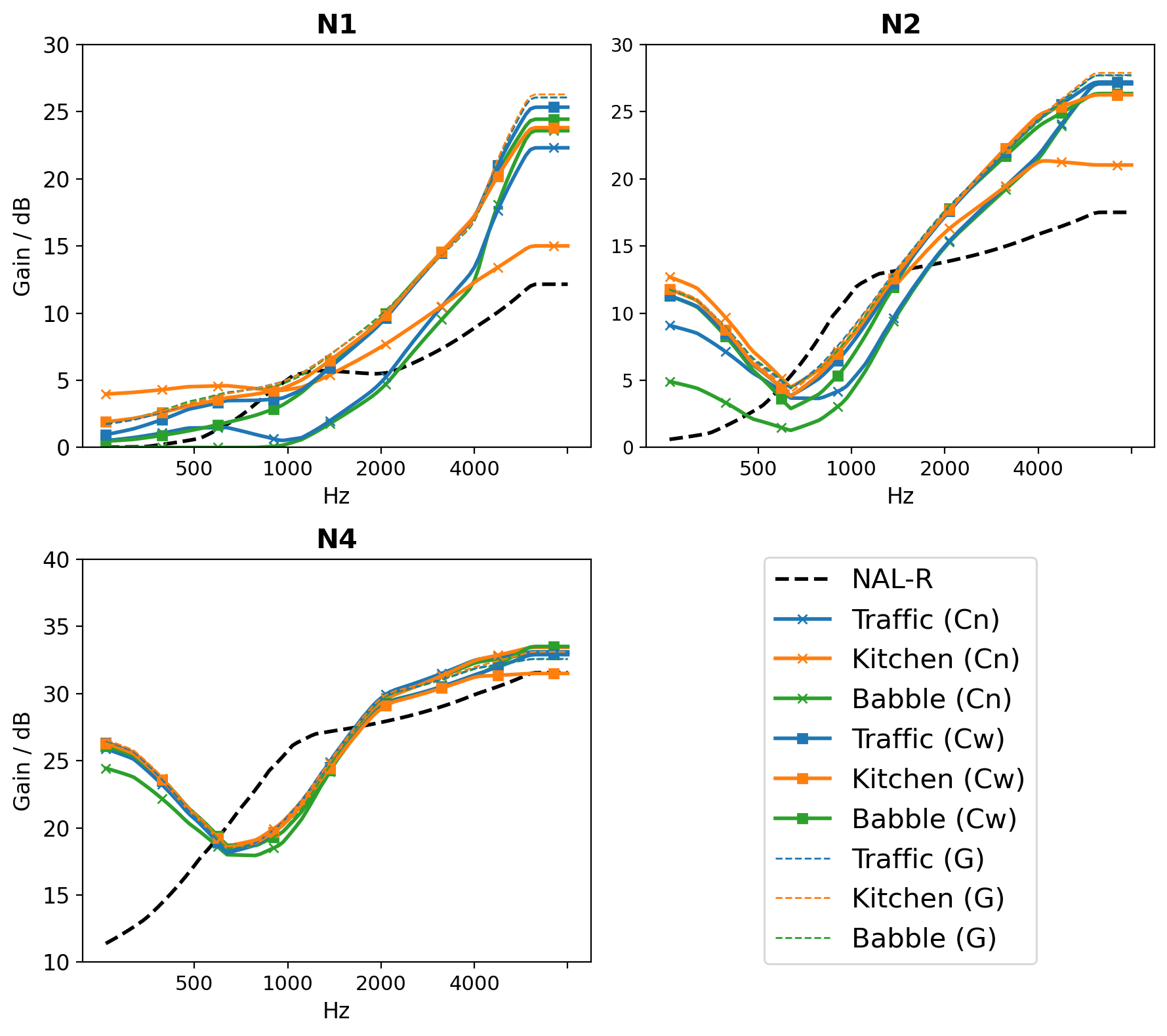}
  \caption{Frequency responses of NAL-R fitting, custom fittings optimised with noisy data (Cn), custom fittings optimised with Wiener filtering enhanced noisy data (Cw), and the general fittings (G) for different hearing losses. }
  \label{fig:FRs}
  \vspace{-.4cm}
\end{figure}

\section{Results and Discussions}
\label{sec:results}

Fig.~\ref{fig:FRs} shows the frequency responses of the optimised general (G) and custom (Cn, Cw) fittings along with the baseline NAL-R prescription for the three hearing loss categories in various noise conditions. First, it can be observed that in general, the frequency responses of optimised fittings have higher gains in low and high frequencies and lower gains around 1\,kHz than the NAR-R prescription. This is broadly consistent with the results of subjective hearing experiments reported in~\cite{mackersie2020hearing} and our previous findings using DHASP~\cite{tu2021dhasp}. The subjective experiments for hearing-aid self-fitting showed that hearing impaired listeners prefer higher gain in high frequency, and lower gain around 1\,kHz for speech in noise compared to the NAL-NL2 fitting, which provides even more gain in low and high frequencies and less gain in middle frequency than NAL-R.

Second, compared to the frequency responses of the processors optimised using noisy data (Cn), those optimised using Wiener filtering denoised data (Cw) are more similar to those optimised using clean data (G). This is expected as Wiener filtering is able to suppress the noise to some extent. Among the three noise types, the kitchen noise energy is mainly distributed in high frequencies, and thus the processors optimised in kitchen noise provide more gain in low frequency and less gain in high frequency. On the contrary, the processors optimised in traffic and babble noises, whose energy is mostly in low frequencies, show more gain in high frequency and less in low frequency. It can also be observed that as the hearing loss severity increases from N1 to N4, the differences among processors optimised with different settings are smaller.

Table~\ref{tab:scores} presents the HASPI, HASQI, and FWSNR scores of the various fittings evaluated in this study. The optimised fittings specific to different listening environments and noise-reduction processing outperform NAL-R in most of the evaluation scores. The improvement over NAL-R is in general larger for more severe hearing loss than that for mild hearing loss, but the benefit can be seen for all three listeners' audiograms and across different noise conditions. This shows there are potential advantages in listening-condition specific fittings that can be learned using the proposed data-driven approach. 

Comparing the scores between the optimised custom fittings and the general prescriptive fittings, it is clear that the custom fittings produced higher FWSNRs than the general fittings in all cases
% , especially for mild and moderate hearing losses. 
The custom fittings also achieved overall higher or approximately equal HASPI and HASQI scores than the general fittings 
for mild and moderate hearing losses.
For moderate to severe hearing loss in the babble noise condition, the general fitting scores are marginally better than the custom fitting scores. This could be due to the fact that the custom fittings provided overall higher insertion gain, and HASPI and HASQI are sensitive to signal presentation level when the hearing threshold is high, i.e., when the hearing loss is more severe. 

It can also be seen that Wiener filtering does not lead to better performance for hearing loss based on the objective evaluation. Our direct measurement without the hearing loss model suggests that Wiener filtering can improve the SNR of noisy speech on average by 6.18, 8.62, and 5.06\,dB for traffic, kitchen and babble noise, respectively. 
%When measured in the FWSNR, the improvement is reduced to 0.10 and 4.05\,dB for traffic and kitchen noise, whereas it decreases the FWSNRs for babble noise by 1.06 dB. 
However, the HASPI scores suggest that only listeners with mild and moderate hearing losses can gain intelligibility benefit from Wiener filtering in the environment with traffic noise, where the noise is relatively stationary and distributed in low frequencies. For more severe hearing loss, or in the kitchen and babble noise, Wiener filtering did not improve the HASPI score. The SNR improvement by Wiener filtering also did not translate into improvement to the FWSNR in all the tested conditions. While it improves the FWSNR for NAL-R fittings in the traffic noise, there is no clear performance pattern for other noise environments. Overall, the objective experiments suggest the potential advantage of the custom hearing aid fitting optimisation, while challenging the intelligibility benefit of Wiener filtering for environmental noise suppression.

% Generally, the optimised fittings can achieve better performance than NAL-R prescription based on the three metrics. Processors optimised with noisy data (N) can always achieve highest FWSNRs, and for mild and moderated hearing losses they could also reach comparatively higher HASPI and HASQI scores in most cases. Processors optimised with clean data (C) can outperform other options for moderate to severe hearing loss according to HASPI and HASQI. The reason could be that they provide overall higher insertion gain and HASPI and HASQI are sensitive to signal presentation level when the hearing threshold is high, i.e. the hearing loss is relatively severe. It can also be noted that when amplification is not applied, Wiener filtering though improves FWSNRs for stationary noise (traffic) and for moderate to severe hearing loss, it does not necessarily help to improve HASPI or HASQI in most cases. Even with amplification, benefit with Wiener filtering can only be observed in HASPI for speech in traffic noise for mild and moderate hearing losses. Overall, these objective results prove the necessity of customised hearing aid fitting optimisation, and challenge the benefit of Wiener filtering in hearing aids.

\section{Conclusions}
\label{sec:conclusions}
This paper has presented a data-driven method to include the effects of different noises and noise suppression techniques into the development of hearing aid fittings. It was argued that a data-driven hearing aid fitting algorithm can be more flexible than current prescribed fitting formulae. A differentiable hearing loss model was proposed for the data-driven optimisation and evaluated in different noise conditions both with and without noise reduction processing enabled. The objective evaluation metrics suggest that the optimisation based approach has potential to outperform prescribed fittings, and that noise-dependent optimisation is particularly promising, with greatest benefits for mild and moderate hearing losses. 
% The differentiable hearing loss model proposed in this paper can be directly embedded into the optimisation of recent powerful deep neural network based methods to help improve the performance of speech enhancement for hearing impaired listeners, similar to the previous work DHASP.

So far the results have been evaluated using objective measurements alone. Based on the promise of these early-stage results, subjective evaluations with hearing impaired listeners are planned for the next phase of the work, including subjective comparisons between the optimisation objectives of the hearing loss model proposed in this work and HASPI based model in the DHASP~\cite{tu2021dhasp}.

\bibliographystyle{IEEEtran}
\bibliography{mybib}

\begin{thebibliography}{10}
\providecommand{\url}[1]{#1}
\csname url@samestyle\endcsname
\providecommand{\newblock}{\relax}
\providecommand{\bibinfo}[2]{#2}
\providecommand{\BIBentrySTDinterwordspacing}{\spaceskip=0pt\relax}
\providecommand{\BIBentryALTinterwordstretchfactor}{4}
\providecommand{\BIBentryALTinterwordspacing}{\spaceskip=\fontdimen2\font plus
\BIBentryALTinterwordstretchfactor\fontdimen3\font minus
  \fontdimen4\font\relax}
\providecommand{\BIBforeignlanguage}[2]{{%
\expandafter\ifx\csname l@#1\endcsname\relax
\typeout{** WARNING: IEEEtran.bst: No hyphenation pattern has been}%
\typeout{** loaded for the language `#1'. Using the pattern for}%
\typeout{** the default language instead.}%
\else
\language=\csname l@#1\endcsname
\fi
#2}}
\providecommand{\BIBdecl}{\relax}
\BIBdecl

\bibitem{lesica2018hearing}
N.~A. Lesica, ``Why do hearing aids fail to restore normal auditory
  perception?'' \emph{Trends in Neurosciences}, vol.~41, no.~4, pp. 174--185,
  2018.

\bibitem{brons2015acoustical}
I.~Brons, R.~Houben, and W.~A. Dreschler, ``Acoustical and perceptual
  comparison of noise reduction and compression in hearing aids,''
  \emph{Journal of Speech, Language, and Hearing Research}, vol.~58, no.~4, pp.
  1363--1376, 2015.

\bibitem{byrne1986national}
D.~Byrne and H.~Dillon, ``The national acoustic laboratories'({NAL}) new
  procedure for selecting the gain and frequency response of a hearing aid,''
  \emph{Ear and Hearing}, vol.~7, no.~4, pp. 257--265, 1986.

\bibitem{keidser2011nal}
G.~Keidser, H.~Dillon, M.~Flax, T.~Ching, and S.~Brewer, ``The nal-nl2
  prescription procedure,'' \emph{Audiology Research}, vol.~1, no.~1, pp.
  88--90, 2011.

\bibitem{moore1997model}
B.~Moore and B.~Glasberg, ``A model of loudness perception applied to cochlear
  hearing loss,'' \emph{Auditory Neuroscience}, vol.~3, no.~3, pp. 289--311,
  1997.

\bibitem{ansi1997s3}
A.~ANSI, ``S3. 5-1997, methods for the calculation of the speech
  intelligibility index,'' \emph{New York: American National Standards
  Institute}, vol.~19, pp. 90--119, 1997.

\bibitem{moore1999use}
B.~Moore, B.~Glasberg, and M.~Stone, ``Use of a loudness model for hearing aid
  fitting: {III}. a general method for deriving initial fittings for hearing
  aids with multi-channel compression,'' \emph{British Journal of Audiology},
  vol.~33, no.~4, pp. 241--258, 1999.

\bibitem{moore2010development}
B.~C. Moore, B.~R. Glasberg, and M.~A. Stone, ``Development of a new method for
  deriving initial fittings for hearing aids with multi-channel compression:
  {CAMEQ2-HF},'' \emph{International Journal of Audiology}, vol.~49, no.~3, pp.
  216--227, 2010.

\bibitem{scollie2005desired}
S.~Scollie, R.~Seewald, L.~Cornelisse, S.~Moodie, M.~Bagatto, D.~Laurnagaray,
  S.~Beaulac, and J.~Pumford, ``The desired sensation level multistage
  input/output algorithm,'' \emph{Trends in Amplification}, vol.~9, no.~4, pp.
  159--197, 2005.

\bibitem{vary2006digital}
P.~Vary and R.~Martin, \emph{Digital speech transmission: Enhancement, coding
  and error concealment}.\hskip 1em plus 0.5em minus 0.4em\relax John Wiley \&
  Sons, 2006.

\bibitem{boll1979suppression}
S.~Boll, ``Suppression of acoustic noise in speech using spectral
  subtraction,'' \emph{IEEE Transactions on Acoustics, Speech, and Signal
  Processing}, vol.~27, no.~2, pp. 113--120, 1979.

\bibitem{bentler2006digital}
R.~Bentler and L.-K. Chiou, ``Digital noise reduction: An overview,''
  \emph{Trends in Amplification}, vol.~10, no.~2, pp. 67--82, 2006.

\bibitem{levitt2001noise}
H.~Levitt, ``Noise reduction in hearing aids: A review,'' \emph{Journal of
  Rehabilitation Research and Development}, vol.~38, no.~1, pp. 111--122, 2001.

\bibitem{lamarche2010adaptive}
L.~Lamarche, C.~Gigu{\`e}re, W.~Gueaieb, T.~Aboulnasr, and H.~Othman,
  ``Adaptive environment classification system for hearing aids,'' \emph{The
  Journal of the Acoustical Society of America}, vol. 127, no.~5, pp.
  3124--3135, 2010.

\bibitem{nordqvist2004efficient}
P.~Nordqvist and A.~Leijon, ``An efficient robust sound classification
  algorithm for hearing aids,'' \emph{The Journal of the Acoustical Society of
  America}, vol. 115, no.~6, pp. 3033--3041, 2004.

\bibitem{tu2021dhasp}
Z.~Tu, N.~Ma, and J.~Barker, ``Dhasp: Differentiable hearing aid speech
  processing,'' in \emph{ICASSP 2021-2021 IEEE International Conference on
  Acoustics, Speech and Signal Processing (ICASSP)}.\hskip 1em plus 0.5em minus
  0.4em\relax IEEE, 2021, pp. 296--300.

\bibitem{baer1993effects}
T.~Baer and B.~C. Moore, ``Effects of spectral smearing on the intelligibility
  of sentences in noise,'' \emph{The Journal of the Acoustical Society of
  America}, vol.~94, no.~3, pp. 1229--1241, 1993.

\bibitem{baer1994effects}
------, ``Effects of spectral smearing on the intelligibility of sentences in
  the presence of interfering speech,'' \emph{The Journal of the Acoustical
  Society of America}, vol.~95, no.~4, pp. 2277--2280, 1994.

\bibitem{moore1993simulation}
B.~C. Moore and B.~R. Glasberg, ``Simulation of the effects of loudness
  recruitment and threshold elevation on the intelligibility of speech in quiet
  and in a background of speech,'' \emph{The Journal of the Acoustical Society
  of America}, vol.~94, no.~4, pp. 2050--2062, 1993.

\bibitem{stone1999tolerable}
M.~A. Stone and B.~C. Moore, ``Tolerable hearing aid delays. i. estimation of
  limits imposed by the auditory path alone using simulated hearing losses,''
  \emph{Ear and Hearing}, vol.~20, no.~3, pp. 182--192, 1999.

\bibitem{graetzer2020clarity}
S.~Graetzer, M.~Akeroyd, J.~P. Barker, T.~J. Cox, J.~F. Culling, G.~Naylor,
  E.~Porter, and R.~V. Munoz, ``Clarity: Machine learning challenges to
  revolutionise hearing device processing,'' \emph{Interspeech}, 2021.

\bibitem{paszke2019pytorch}
A.~Paszke, S.~Gross, F.~Massa, A.~Lerer, J.~Bradbury, G.~Chanan, T.~Killeen,
  Z.~Lin, N.~Gimelshein, L.~Antiga \emph{et~al.}, ``Pytorch: An imperative
  style, high-performance deep learning library,'' \emph{Advances in Neural
  Information Processing Systems}, vol.~32, pp. 8026--8037, 2019.

\bibitem{shaw1974transformation}
E.~A. Shaw, ``Transformation of sound pressure level from the free field to the
  eardrum in the horizontal plane,'' \emph{The Journal of the Acoustical
  Society of America}, vol.~56, no.~6, pp. 1848--1861, 1974.

\bibitem{killion1978revised}
M.~C. Killion, ``Revised estimate of minimum audible pressure: Where is
  the’’missing 6 db’’?'' \emph{The Journal of the Acoustical Society of
  America}, vol.~63, no.~5, pp. 1501--1508, 1978.

\bibitem{moore1983suggested}
B.~C. Moore and B.~R. Glasberg, ``Suggested formulae for calculating
  auditory-filter bandwidths and excitation patterns,'' \emph{The Journal of
  the Acoustical Society of America}, vol.~74, no.~3, pp. 750--753, 1983.

\bibitem{glasberg1990derivation}
B.~R. Glasberg and B.~C. Moore, ``Derivation of auditory filter shapes from
  notched-noise data,'' \emph{Hearing Research}, vol.~47, no. 1-2, pp.
  103--138, 1990.

\bibitem{kates2014hearinghaspi}
J.~M. Kates and K.~H. Arehart, ``The hearing-aid speech perception index
  (haspi),'' \emph{Speech Communication}, vol.~65, pp. 75--93, 2014.

\bibitem{kates2014hearinghasqi}
------, ``The hearing-aid speech quality index (hasqi) version 2,''
  \emph{Journal of the Audio Engineering Society}, vol.~62, no.~3, pp. 99--117,
  2014.

\bibitem{loizou2013speech}
P.~C. Loizou, \emph{Speech Enhancement: Theory and Practice}.\hskip 1em plus
  0.5em minus 0.4em\relax CRC press, 2013.

\bibitem{bisgaard2010standard}
N.~Bisgaard, M.~S. Vlaming, and M.~Dahlquist, ``Standard audiograms for the iec
  60118-15 measurement procedure,'' \emph{Trends in Amplification}, vol.~14,
  no.~2, pp. 113--120, 2010.

\bibitem{valentini2016investigating}
C.~Valentini-Botinhao, X.~Wang, S.~Takaki, and J.~Yamagishi, ``Investigating
  rnn-based speech enhancement methods for noise-robust text-to-speech.'' in
  \emph{SSW}, 2016, pp. 146--152.

\bibitem{veaux2013voice}
C.~Veaux, J.~Yamagishi, and S.~King, ``The voice bank corpus: Design,
  collection and data analysis of a large regional accent speech database,'' in
  \emph{2013 international conference oriental COCOSDA held jointly with 2013
  conference on Asian spoken language research and evaluation
  (O-COCOSDA/CASLRE)}.\hskip 1em plus 0.5em minus 0.4em\relax IEEE, 2013, pp.
  1--4.

\bibitem{thiemann2013demand}
J.~Thiemann, N.~Ito, and E.~Vincent, ``Demand: a collection of multi-channel
  recordings of acoustic noise in diverse environments,'' in \emph{Proc.
  Meetings Acoust}, 2013.

\bibitem{itu1993objective}
P.~ITU-T, ``Objective measurement of active speech level,'' \emph{ITU-T
  Recommendation}, 1993.

\bibitem{plapous2006improved}
C.~Plapous, C.~Marro, and P.~Scalart, ``Improved signal-to-noise ratio
  estimation for speech enhancement,'' \emph{IEEE Transactions on Audio,
  Speech, and Language Processing}, vol.~14, no.~6, pp. 2098--2108, 2006.

\bibitem{mackersie2020hearing}
C.~L. Mackersie, A.~Boothroyd, and H.~Garudadri, ``Hearing aid self-adjustment:
  Effects of formal speech-perception test and noise,'' \emph{Trends in
  Hearing}, vol.~24, 2020.

\end{thebibliography}
\end{document}